 \documentclass[final,5p,times,twocolumn,authoryear]{elsarticle}

\usepackage{amssymb}
\usepackage{lipsum}
\usepackage{amsthm}
\usepackage{xcolor}
\usepackage{amsmath}
\usepackage{comment}

\def\msun{M$_\odot$}

\journal{Physics Letters B}

\begin{document}

\begin{frontmatter}



\title{Towards an anomaly detection pipeline for gravitational waves at the Einstein Telescope}


\author[first]{Gianluca Inguglia}
\author[first]{Huw Haigh}
\author[first]{Kristyna Vitulova}
\author[first]{Ulyana Dupletsa}

\affiliation[first]{organization={Austrian Academy of Sciences, Marietta Blau Institute for Particle Physics  (MBI)},
            city={Vienna},
            postcode={1010}, 
            country={Austria}}


\begin{abstract}
We present the implementation of an anomaly–detection algorithm based on a deep convolutional autoencoder for the search for gravitational waves (GWs) in time-frequency spectrograms. Our method targets short-duration ($\lesssim 2\,\text{s}$) GW signals, exemplified by mergers of compact objects forming or involving an intermediate-mass black hole (IMBH). Such short signals are difficult to distinguish from background noise; yet their brevity makes them well-suited to machine-learning analyses with modest computational requirements. 
Using the data from the Einstein Telescope Mock Data Challenge as a benchmark, we demonstrate that the approach can successfully 
flag GW-like transients as anomalies in interferometer data of a single detector, achieving an initial detection efficiency of 23\% for injected signals corresponding to IMBH–forming mergers. 
After introducing weak supervision, the model exhibits excellent generalisation and recovers all injected IMBH–forming mergers, independent of their total mass or signal–to–noise ratio, with a false–alarm rate due to statistical noise fluctuations of approximately 4.5 events per year for a single interferometer operating with a 100\% duty cycle. The method also successfully identifies lower–mass mergers leading to the formation of black holes with mass larger than $\simeq20\,M_\odot$. 
Our pipeline does not yet classify anomalies, distinguishing between actual GW signals and noise artefacts; however, it highlights any deviation from the learned background noise distribution for further scrutiny. These results demonstrate that anomaly detection offers a powerful, model–independent framework for future GW searches, paving the way toward fully automated and adaptive analysis pipelines.
\end{abstract}




\end{frontmatter}




\section{Introduction}
\label{sec:intro}

The LIGO-Virgo-KAGRA (LVK) Collaboration currently employs both modeled and unmodeled search techniques for gravitational-wave (GW) detection (see~\cite{LIGOScientific:2025yae} for a recent overview), with a preference for modeled searches based on matched filtering. In this approach, precomputed waveform templates representing compact binary coalescences (CBCs) form a template bank that is cross-correlated with the interferometer data. When a template closely matches a potential signal, a trigger is generated, indicating a candidate GW transient for further inspection and validation.

Although highly effective in signal detection, matched filtering is limited by the size and computational cost of the template bank. Since only a limited number of waveform templates can be included in the template bank, mismatches - a measure of how well a template approximates a true signal - are inevitable. It is standard practice to accept mismatches of up to 3\% between a real signal and a template~\citep{Sakon:2022ibh}, corresponding to a reduction in the signal-to-noise ratio (SNR) of up to 1.5\%. This seemingly small SNR loss could be critical, especially for signals near the detection threshold, which might be missed.

Unmodeled searches, on the other hand, aim to reconstruct GW signals without relying on specific waveform models. They target a wide variety of sources, and especially short, burst-like signals, which are the most difficult to distinguish from background noise and instrumental artefacts. Such short-duration, hard-to-model signals are ideal targets for machine learning algorithms, which can learn statistical patterns and representations directly from data. Machine and deep learning provide a powerful framework for classification, regression, and inference tasks, and are now widely adopted across most scientific domains. In deep learning, this is typically achieved through multi-layer neural networks trained via gradient-based optimisation to minimise task-specific loss functions~\citep{Cuoco:2020ogp,Albertsson:2018maf,Feichtinger:2021uff,Yan:2025hfq,Schafer:2022dxv}.

Exploring machine learning approaches is becoming increasingly crucial in light of third-generation GW interferometers such as the Einstein Telescope (ET)~\citep{Punturo:2010zz, Hild:2010id, Branchesi:2023mws, Abac:2025saz} or Cosmic Explorer (CE)~\citep{Reitze:2019iox, Evans:2021gyd, Gupta:2023lga}, representing the European and the United States proposals, respectively, for future ground-based GW detectors. Their improved sensitivity across a broader frequency range will lead to a substantial increase in detectable events, resulting in higher computational demands, not only for simulated data generation but also for detection and analysis of real data. This motivates the development of complementary or alternative search methods to enhance the discovery potential of current and future GW observatories.

We reformulate GW searches as a general anomaly-detection problem: instead of matching data to predefined templates, the algorithm learns the structure of detector noise in frequency–time spectrograms and flags significant deviations as potential signals. In this work, we present the implementation and evaluation of an anomaly-detection algorithm for searching GW transients, based on convolutional autoencoders operating on spectrogram data from a single ET interferometer\footnote{As detailed in Sec.~\ref{sec:results}, for single ET interferometer, we mean that we consider only one out of the 3 detector channels}. We focus on burst signals, short GW transients that are produced, for instance, by the mergers of two black holes (BHs) involving or resulting in the formation of an intermediate-mass black hole (IMBH), a class of BHs with masses in the range  $10^2-10^5$\,\msun~\citep{Greene_2020}. We then demonstrate that our model is scalable and applicable across a broader range of BH masses.

The paper is structured as follows. In Sec.~\ref{sec:IMBH} we introduce the science case of interest. We detail the network architecture and definition of a simple loss algorithm in Sec.~\ref{sec:CAE}. In Sec.~\ref{sec:results}, we then evaluate the ability of the model to reconstruct ET noise-only spectrograms and define the criteria by which observed data may be flagged as anomalous. We then investigate the performance of the model using the first Einstein Telescope Mock Data Challenge (MDC)~\citep{Tania:2025bsa,Regimbau:2012ir} dataset, focusing specifically on BBH mergers where the sum of source frame masses is greater than 100 M$_\odot$. We conclude our discussion in Sec.~\ref{sec:conclusions}, highlighting future directions of this work.

\section{Intermediate-mass black holes and their gravitational wave signatures}\label{sec:IMBH}
Up to date, the LVK Collaboration has observed 218 GW candidate events, mostly associated with the merger of two black holes, as released in the latest GWTC-4.0 catalog~\citep{LIGOScientific:2025slb}. Remarkably, several of these mergers involved IMBHs either in the merging components, as in the case of GW190521~\citep{LIGOScientific:2020iuh, LIGOScientific:2020ufj} or in the remnant, as for GW231123~\citep{LIGOScientific:2025rsn}.

\begin{figure}[!ht]
\begin{center}
\resizebox{9cm}{!}{
\includegraphics{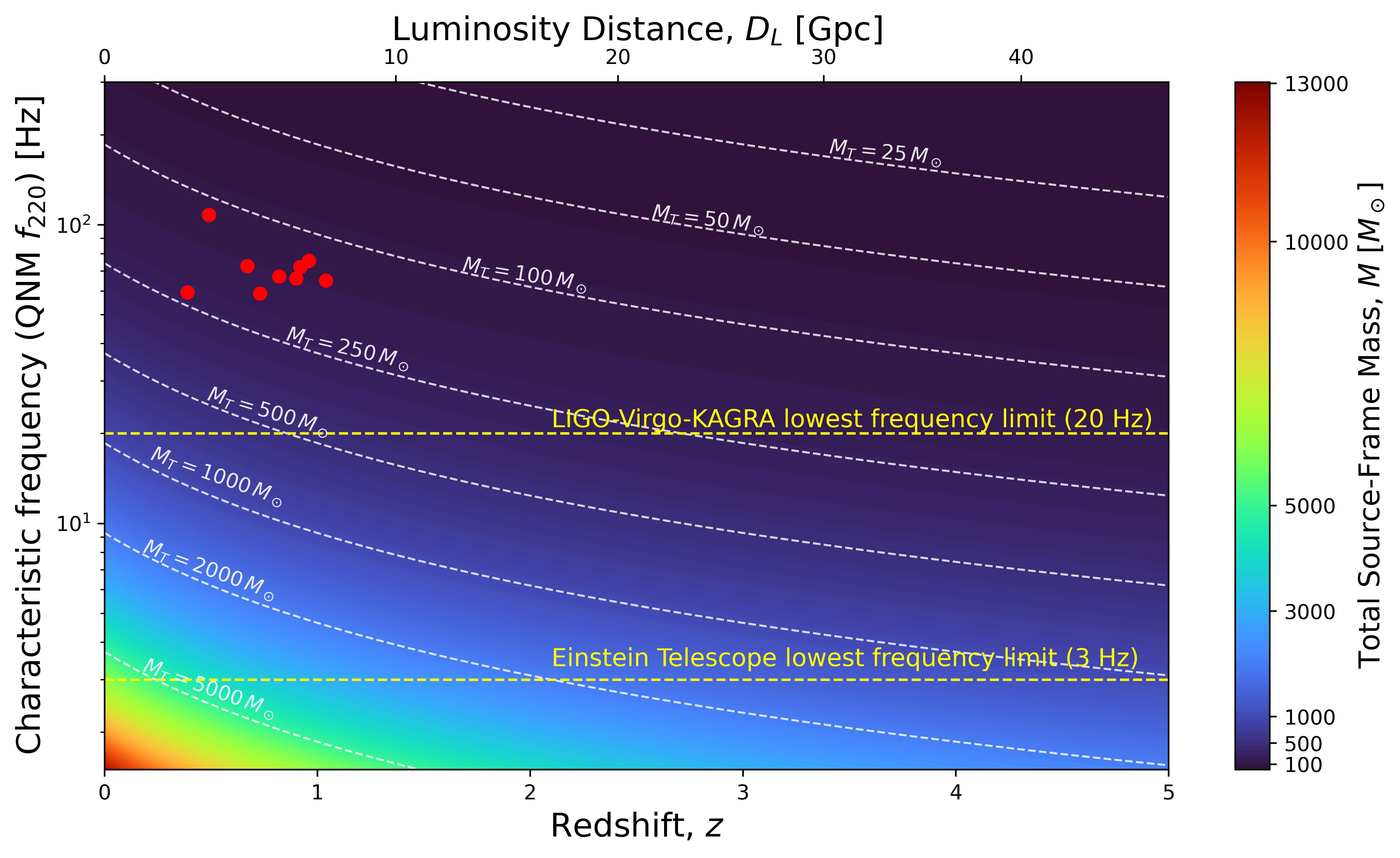}
}\caption{Theoretical frequency $f_{220}$ as a function of redshift for binary black hole mergers assuming different total source-frame masses~\citep{Berti:2005ys}, calculated using $\alpha(a_f) \simeq 0.536$ and $\varepsilon = 0.07$~\citep{Keitel:2016krm,Healy:2014eua}. See ~\ref{app:qnm_freq} for further details. The red points indicate IMBH candidates detected via GWs, including GW190521~\citep{LIGOScientific:2020iuh, LIGOScientific:2020ufj} and GW231123~\citep{LIGOScientific:2025rsn}. The yellow dashed lines indicate the lowest frequency limit of the LIGO-Virgo-KAGRA detectors network and of the Einstein Telescope~\citep{Abbott:2020lrr, Abac:2025saz}.}\label{fig:merger}
\end{center}
\end{figure}

One critical aspect of the GWs emitted during the formation of an IMBH, or during a merger involving IMBHs, is that the characteristic strain amplitude, for this type of transients, is distributed towards lower frequencies, well below 100 Hz, decreasing when the redshifted mass, $M_z=(1+z)M(=m_1+m_2)$, increases. 
Figure~\ref{fig:merger} shows the evolution of the merger frequency, exemplified by the $f_{220}$ mode, for several total source-frame masses as a function of the redshift, indicating that the heavier and the further away the events, the lower the merger frequency, reaching values below the sensitivity level of next-generation ground-based interferometers such as ET.

Second-generation detectors such as LIGO or Virgo are only sensitive to these types of events when the total mass is of the order of up to a few hundred $M_\odot$, with the caveat that only the final merger phase can be detected as a "burst" due to the inspiral phase happening mostly at frequencies below the sensitivities of the detectors. In the case of GW190521 and GW231123 only 4 and 5 cycles were observed in the frequency band of 30-80 Hz, respectively, as reported in ~\citep{LIGOScientific:2020iuh, LIGOScientific:2020ufj} and in ~\citep{LIGOScientific:2025rsn}; with higher total source-frame masses less cycles will be in the sensitive region of the detector, making these events more difficult to be separated from environmental or instrumental transients (or glitches). ET is designed to increase the sensitivity to GW strains of at least an order of magnitude with respect to second-generation detectors, and will therefore be able to capture GW signals happening at lower frequencies, also thanks to its cryogenic operations, which in turn will allow us to extend to higher total masses the sensitivity of the detector. In ET, events like GW190521 or GW231123 will be detected with the final phase of the inspiral, typically staying within the detector-sensitive region for up to a few seconds. For events with total masses up to 0.5-1$\times 10^4$ M$_\odot$, however, only a few cycles of the final merger phase will be detected by ET, and more massive events will fall outside the sensitivity of ET. Interestingly, space-based observatories planned to be operating at the same time of ET or CE, as LISA, will be able to capture the inspiral phase of such massive mergers, with the possibility of multi-band detections~\citep{Abac:2025saz}. 

Because mergers involving or resulting in the formation of an IMBH radiate GWs primarily at low frequencies, the development of a detection algorithm capable of identifying such signals, irrespective of their morphology or distance and even within single-detector data, would be of great interest for both current and future GW observatories.
\section{Autoencoders as anomaly detection tools}\label{sec:CAE}
Autoencoders (AEs) are a class of unsupervised learning algorithms in which input data are encoded into a generally low-dimensional \textit{latent space} and subsequently decoded back into the original data domain~\citep{2020arXiv200305991B}. Through this process, the AE learns a compressed internal representation that captures the most essential features of the input. The quality of the reconstruction after encoding and decoding is quantified by the \textit{reconstruction error}, typically measured using the mean-squared-error (MSE) loss function:
\begin{equation}
\mathcal{L}_{\mathrm{MSE}} = \frac{1}{N} \sum_{i=1}^{N} (x_i - \hat{x}_i)^2,
\end{equation}
where \(x_i\) and \(\hat{x}_i\) denote the input and its reconstructed counterpart, respectively. 

For each input sample \(x^{(j)}\), the reconstruction error quantifies the discrepancy between the input and its reconstruction \(\hat{x}^{(j)}\), and is defined as
\begin{equation}
E_{\mathrm{rec}}^{(j)} = \frac{1}{N} \sum_{i=1}^{N} \left(x_i^{(j)} - \hat{x}_i^{(j)}\right)^2,
\end{equation}
where \(N\) denotes the number of input features (e.g., pixels or time–frequency bins). During training, the autoencoder minimises the average reconstruction error across a batch of size \(B\),
\begin{equation}
\mathcal{L}_{\mathrm{MSE}} = \frac{1}{B} \sum_{j=1}^{B} E_{\mathrm{rec}}^{(j)},
\end{equation}
which serves as the loss function. After training, individual values of \(E_{\mathrm{rec}}^{(j)}\) can be used to identify inputs that deviate from the learned data distribution, such as potential anomalies. A small value of \(E_{\mathrm{rec}}^{(j)}\) indicates that the autoencoder successfully reproduces the input. In contrast, a large value suggests a deviation from the learned data manifold, making AEs particularly suitable for anomaly detection. A typical example is the presence of a signal embedded in noise. When an AE is trained exclusively on noise-only data (for example, time series or spectrograms), it learns to reconstruct this noise accurately. However, when presented with data containing both noise and a signal, the AE fails to reconstruct the signal component, resulting in a noticeably larger reconstruction error.

By building a reference distribution of reconstruction errors from the noise-only training set, one can define a threshold above which test data are flagged as anomalous. This approach enables the detection of subtle effects in (high-dimensional) datasets without the explicit need for labeled signal examples.\\In the context of GW searches, this aspect can have a significant impact, as the method does not rely on the information about the morphology of the signal waveform.\\Several variants of autoencoders have been developed to adapt the basic architecture to different types of data and tasks. Convolutional autoencoders (CAEs) are well-suited for structured inputs such as images or spectrograms, as they employ convolutional layers to extract spatially localised features~\citep{Morawski:2021kxv,Pol:2018nhq,Gandrakota:2024yqs, Bacon:2022lsm}. Variational autoencoders (VAEs)~\citep{2019arXiv190602691K, Oshino:2025laz} introduce a probabilistic framework by learning a distribution over the latent space, enabling generative modelling and smooth interpolation between data points. Other extensions include denoising autoencoders (DAEs)~\citep{2013arXiv1305.6663B}, which learn to reconstruct clean inputs from noisy versions, and sparse autoencoders (SAEs)~\citep{2023arXiv230908600C}, which promote sparsity in the latent representation. 

\begin{figure}[!ht]
\begin{center}
\resizebox{9cm}{!}{
\includegraphics{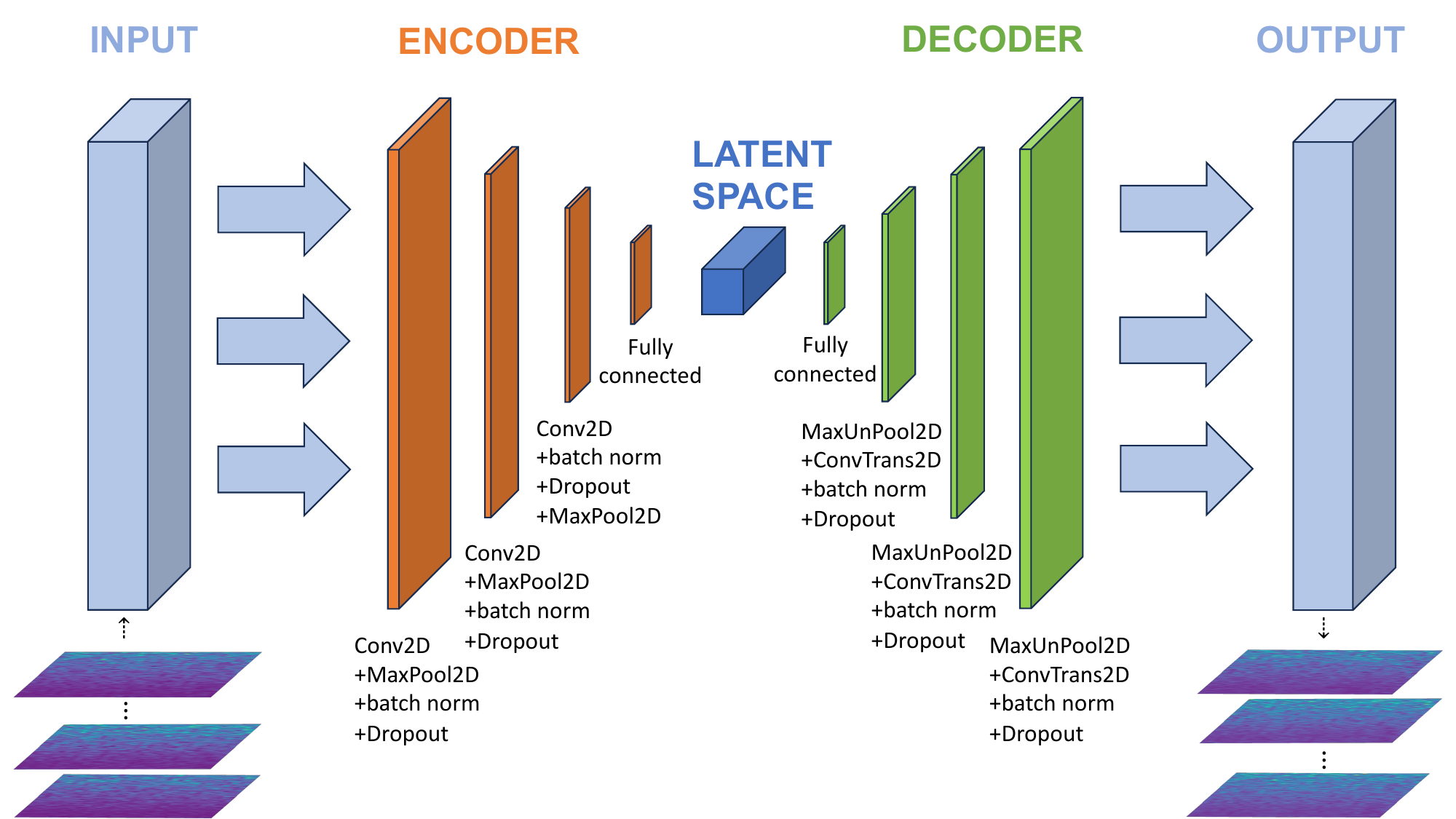}
}\caption{Schematic of the convolutional autoencoder (CAE) architecture adopted for anomaly detection of gravitational waves. The input data — spectrograms of two-second segments of noise strain — are encoded into a latent representation through three convolutional layers, and subsequently decoded via mirrored deconvolutional layers to reconstruct the input. The CAE is trained on noise-only data, enabling it to detect anomalous spectrograms (e.g., containing GW transients) through deviations in reconstruction error.}\label{fig:CAE}
\end{center}
\end{figure}

The choice of autoencoder type depends on the structure of the input data and the intended application. In our case, we adopt CAEs because our input data consists of images, specifically spectrograms (time vs. frequency), which contain noise. We opt for a minimal architecture to minimise computational cost and training time, as shown in Fig.~\ref{fig:CAE}. We implemented a deep CAE in PyTorch~\citep{Paszke:2019xhz} to learn a compact representation of spectrogram images and perform anomaly detection. The model consists of three convolutional encoding blocks followed by three symmetric decoding blocks; each encoding block comprises a convolutional layer (with kernel size 3 and padding 1), batch normalisation, dropout to prevent the model from focusing on some specific features or neurons overfitting the data, and max pooling (with indices retained for unpooling). The decoder mirrors the encoder using max unpooling layers, followed by transposed convolutions, batch normalisation, and dropout, to reconstruct the input. 
The architecture compresses the input through the convolutional layers and learns to reconstruct the original spectrogram by minimising the MSE between input and output. The input spectrograms were grayscale images of size $256 \times 31$ pixels, corresponding to 256 frequency bins and 31 time steps. Initial development of the network architecture was conducted using simple white-noise spectrograms (see \ref{app:implementation}). This allows for faster trial-and-error testing before moving on to realistic noise samples. The encoder applies three successive convolutional layers, each followed by max pooling, halving the spatial resolution at each stage. The spatial dimensions of the feature maps evolve as follows:
\[
256 \times 31 \rightarrow 128 \times 15 \rightarrow 64 \times 7 \rightarrow 32 \times 3.
\]
The number of feature maps increases across layers: 32 after the first convolution, 64 after the second, and 128 after the third, with $12{,}288$ corresponding features. A fully connected layer is then applied at the bottleneck to reduce the latent space dimensionality to 1024. The decoder reconstructs the input resolution using max unpooling guided by the stored indices and transposed convolutions. 

The model was trained for 60 epochs using the Adam optimiser with a learning rate of $2 \times 10^{-3}$ and a dropout factor of 0.08. We used a ReduceLROnPlateau learning rate scheduler (factor 0.1, patience 3 epochs). The training is done using ~120,000 noise-only spectrograms from the ET mock data challenge set. The data are split 70\%/30\% into training/validation sets, with the latter used to track validation loss during training. The loss is defined as the MSE between the input and the reconstructed output. 

Training and validation were performed on mini-batches using GPU acceleration. We use a local dedicated machine for training, validation, testing and further optimisation configured with $2\times16-$core CPU at 4.1 GHz, and $4\times$ Nvidia RTX 400 Ada with 20 GB dedicated VRAM each for hardware acceleration of which one only utilised for training and validation (with a computational time for either supervised or weakly supervised training/validation of less than 1 hour).

\subsection{Introduction of weak supervision}


To further improve the separation in reconstruction loss between noise-only and signal-injected spectrograms, we introduce a secondary training dataset containing anomalous spectrograms. These are created by injecting gravitational wave signals, originating from mergers of varied mass pairings and distances, into the ET noise spectrograms. We inject signals with source frame masses in the range [80, 150] and luminosity distances ranging from 10 to 40 Gpc.

We then define a loss function, still based on the MSE loss, that seeks to penalise the model for reconstructing such anomalous spectrograms with similar MSE error as it does noise-only spectrograms. This separation in reconstruction is captured using the rectified linear unit (ReLU) activation function, which returns 0 for input negative values and returns the input otherwise: 

\begin{equation}
    \Delta \mathcal{L} = \text{ReLU}(m - (\mathcal{L}_{anom} - \mathcal{L}_{noise}))
\end{equation}
where $\mathcal{L}_{anom}, \mathcal{L}_{noise}$ are the reconstruction MSE of signal-injected and noise-only spectrograms, respectively, and $m$ is an introduced hyperparameter that quantifies the desired level of separation in reconstruction error. This function then returns 0 when a level of separation equal to or greater than $m$ is reached:

\begin{equation}
\Delta \mathcal{L} =
\begin{cases}
m - (\mathcal{L}_{anom} - \mathcal{L}_{noise}), & \text{if } \mathcal{L}_{anom} - \mathcal{L}_{noise} < m, \\
0, & \text{if } \mathcal{L}_{anom} - \mathcal{L}_{noise} \ge m.
\end{cases}
\end{equation}

The final loss can then be defined as the sum of the regular MSE loss of noise-only spectrograms and this separation term. As such, the calculated loss is therefore only affected by the reconstruction error associated with anomalous spectrograms in such cases where they are not well separated from the noise-only spectrograms. With adequate separation, the standard MSE loss of the noise-only spectrograms is recovered. 

\begin{equation}
 \mathcal{L}_{total} = \mathcal{L}_{MSE}(\hat{x}_{noise}) + \Delta \mathcal{L}
\end{equation}
We introduce the use of Optuna in optimisation of the model hyperparameters~\citep{akiba2019optuna}, including the newly defined separation margin, $m$. A value of $m=0.05$ was found to induce strong separation between noise-only and signal-injected spectrograms, while maintaining stability of the training process. Higher values of $m$ were found to introduce instability into the results. We notice that in other fields, such as high-energy physics or collider phenomenology, a similar approach to optimising losses but with different function choices was implemented for example in~\cite{Khosa:2020qrz,Banda:2025nrv,Park:2020pak}. 

\section{Noise modelling and signal injection}
\label{sec:results}
This section is divided into two sub-sections. The first of these is dedicated to the analysis of the noise-only data and how the autoencoder handles the associated spectrograms. In the latter sub-section, we then investigate the results when signals are injected into the noise. This is done using BBH merger data provided in the MDC1 challenge; we focus on those mergers involving, or resulting in IMBHs, but also study the ability of the model to generalise to other waveform signatures. 

\subsection{Noise modelling}

This study utilises the Einstein Telescope mock data challenge dataset, which corresponds to approximately one month of simulated observations~\citep{Tania:2025bsa}. This dataset is provided as a time series sampled at 8192 Hz, segmented into 2048-second chunks, and includes realistic noise strains for the three interferometers $E_1$, $E_2$, and $E_3$ in ET’s triangular configuration. Since the final location of the detector is not known yet, the mock data were prepared by positioning ET on the Virgo site in Cascina, with the $E_1$ vertex coincident with the vertex of Virgo as described in~\citep{Tania:2025bsa,Regimbau:2012ir}. 

\begin{figure}[!ht]
\begin{center}
\resizebox{9cm}{!}{
\begin{tabular}{c}
\includegraphics{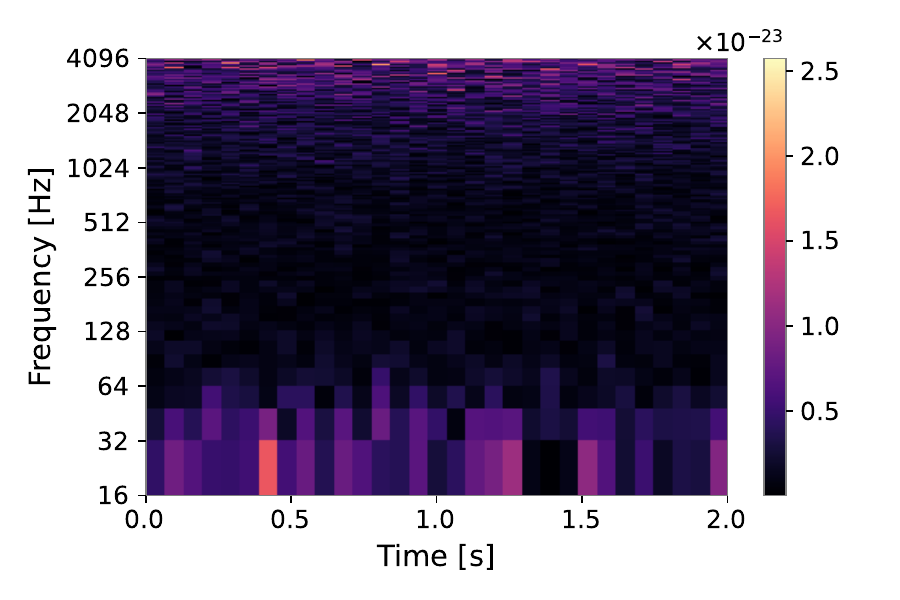}
\end{tabular}}
\caption{An example of the ET noise-only spectrograms used in the training and testing of the convolutional autoencoder. Note that the y-axis (frequency) is scaled by log$_2$ to highlight the lower frequency region that is particularly of interest when searching for IMBH mergers. For illustrative purposes, this spectrogram has not been whitened or normalised.}
\label{fig:ET_noise_spectrogram}
\end{center}
\end{figure}

For simplicity, this study does not yet implement the antenna patterns of the detector and restricts analysis to only the $+$ polarisation component. We limit ourselves to selecting only one detector channel (E1), noting that the choice among any of the three detectors has a negligible impact. We estimate the power spectral density (PSD) using Welch's method (an averaged FFT-based estimator) and whiten the data accordingly. The whitened time series is then split into 2-second segments to construct spectrograms. An example of such is shown in fig. \ref{fig:ET_noise_spectrogram}. 

Before testing the model's ability to recover GW signals amongst background noise, it is important first to understand how it handles pure noise. With a comprehensive knowledge of the expected MSE loss outputs from such noise-only spectrograms, we then define a threshold value, above which a spectrogram is flagged as anomalous and therefore deserving of further analysis. We define this threshold, based on the expected statistical distribution of noise-only spectrograms, as:
\[
\text{Threshold} = \mu + n\sigma,
\]
where $\mu$ and $\sigma$ are the mean and standard deviation of the reconstruction errors in the noise-only test set. One can choose $n$ to suit a desired minimum level of statistical confidence when flagging a spectrogram as anomalous. This choice is made under the assumption of uncorrelated individual $2\,\mathrm{s}$ segments and one-sided Gaussian tail statistics. We can then estimate an associated (per year) false--alarm rate (FAR) with: 
\[
\mathrm{FAR} \simeq \frac{T_{\mathrm{year}}}{2\,\mathrm{s}} \, p 
   = \frac{3.1536\times10^{7}}{2} \, p,
\]
where $p = 1 - \Phi(n\sigma)$ is the one--sided tail probability of a standard normal distribution. Hence, assuming a hypothetical experimental duty cycle of 100\% over a whole year, for thresholds of
\[
\begin{aligned}
3\sigma &\;\Rightarrow\; p \simeq 1.35\times10^{-3}
    \;\Rightarrow\; \mathrm{FAR} \simeq 2.1\times10^{4}\ \mathrm{events/yr},\\[4pt]
5\sigma &\;\Rightarrow\; p \simeq 2.87\times10^{-7}
    \;\Rightarrow\; \mathrm{FAR} \simeq 4.5\ \mathrm{events/yr}.
\end{aligned}
\]

\begin{figure}[!ht]
\begin{center}
\resizebox{9cm}{!}{
\begin{tabular}{c}
\includegraphics{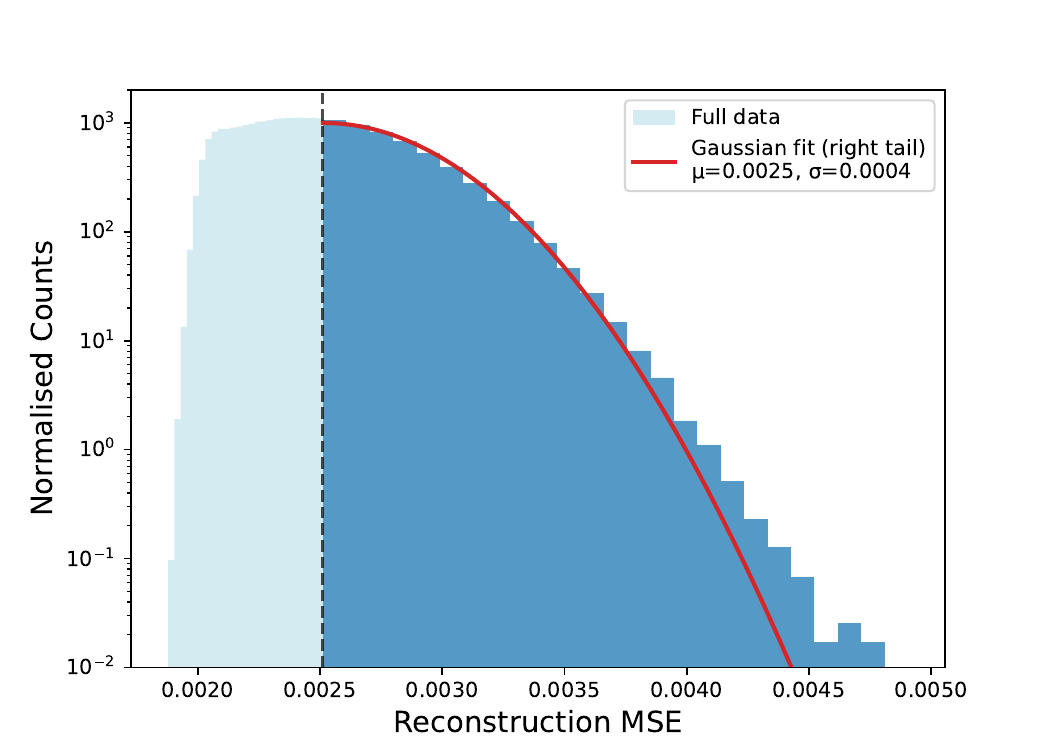}
\end{tabular}}
\caption{The distribution of reconstruction MSE found in all spectrograms in the ET noise samples. The distribution is shown in logarithmic scale with the fitted Gaussian distribution shown in red. The distribution begins to deviate from the Gaussian modelling in the far extreme of the tail; in this region the data were instead fit with a generalised pareto function in order to extract a 5$\sigma$ threshold.}
\label{fig:tail_modelling}
\end{center}
\end{figure}
This treatment can be further extended to include multiple (non-local) detectors with uncorrelated noise, in which case the probability of false alarm is therefore the product of the individual values. For just two detectors, this would further lower the per-year FAR to 29 and 1.3$\times 10 ^{-6}$ with 3 and 5$\sigma$ thresholds, respectively. Such a prediction is, of course, solely concerned with purely statistical fluctuation of noise-only measurements. In future works, one could further include the impact of glitches and other additional sources of uncertainty or false--alarms in the data.
We note that at extremes of the reconstruction MSE distribution, the assumption of Gaussian modelling appears to break down, with a slightly heavier tail extending beyond expected ranges, stemming from particularly signal-like noise spectrograms. Figure \ref{fig:tail_modelling} shows in blue the distribution of reconstruction MSE from spectrograms spanning the whole 1--month MDC dataset. This is shown in log scale with a fitted Gaussian depicted by a red line. The distribution of MSE values exhibits Gaussian-like properties in the upper tail only, with the lower tail (shown in faded blue) dropping sharply instead. We therefore fit the Gaussian using only the upper side (shown in solid blue). This models the distribution well up to an approximate MSE of 0.0038.

\begin{figure}[!ht]
\begin{center}
\resizebox{9cm}{!}{
\begin{tabular}{c}
\includegraphics{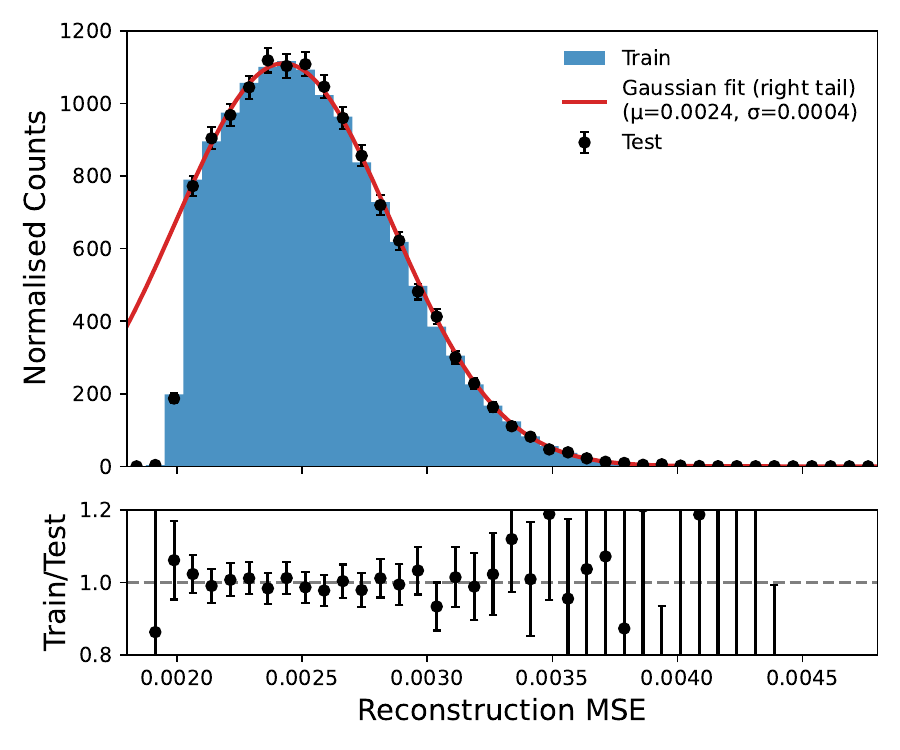}
\end{tabular}}
\caption{Distributions of the reconstruction MSE from the training (blue) and testing (black points) ET noise-only spectrograms. The ratio of the two distributions is shown in the bottoms section, the flat distribution indicates that the model generalises well to unseen input data. }
\label{fig:train_test}
\end{center}
\end{figure}

While this leaves the 3$\sigma$ threshold modelling largely unaffected, it does introduce the need for careful derivation of the higher 5$\sigma$ threshold to avoid underestimation and unexpected false--alarms. We make use of the \textit{generalised pareto} function to model the extrema of the tail. The peak-over-threshold method (POT) is then used to provide an estimate for the 5$\sigma$ threshold. The 3 and 5$\sigma$ thresholds are defined at MSE values of 0.0038 and 0.005, respectively. It should be noted that in this work, we limit the study of false--alarm rates associated with the model to just these threshold values, and do not assign any estimated FAR to individual mergers from the MDC dataset. In future work, this treatment could be extended to calculate a naive FAR associated with a given anomaly based on its reconstruction MSE.
As with any implementation of machine learning methods, it is essential to ensure that the given model is not \textit{over-trained}, whereby it learns features specific only to the training dataset that are not present in unseen data. Over-training leads to unexpected behavior when the model is applied to real-world data and would thereby negate any results. As a check for such behavior we compare the distribution of reconstruction MSE values from the training ET noise and a separate test sample that has not been used in training. Figure \ref{fig:train_test} shows the training distribution in blue and the test sample with black points. The ratio of these two distributions is given in the lower section, with error bars showing the statistical uncertainty associated with each distribution. The distributions show good agreement, thus indicating that the model generalises well to unseen input data.

\subsection{Signal injection}

\begin{figure}[!ht]
\begin{center}
\resizebox{9cm}{!}{
\includegraphics[width=0.5\textwidth,trim={0 15mm 0 25mm},clip]{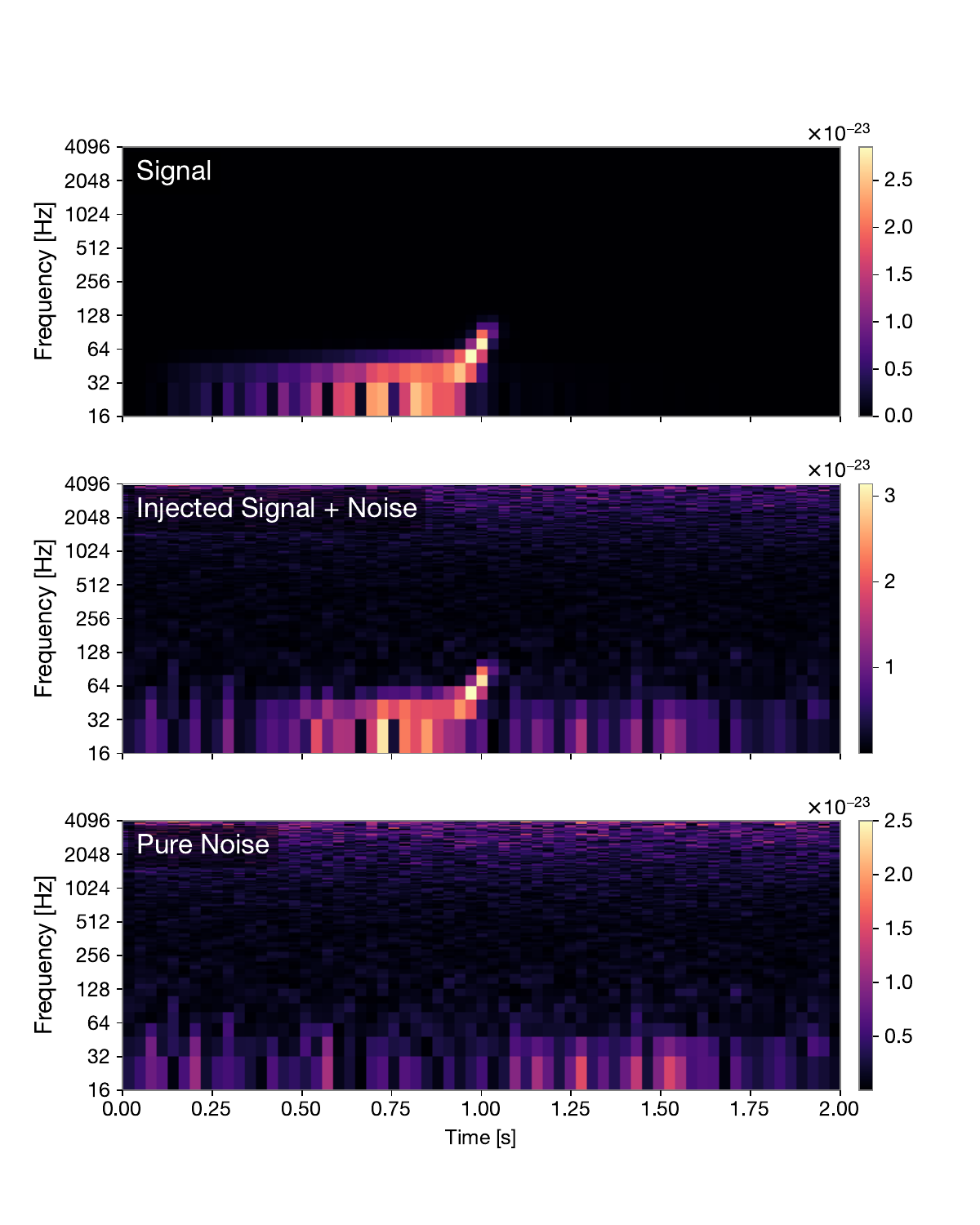}
}
\caption{Example distributions used in the training and testing of the CAE. The lower panel shows a spectrogram of the Einstein Telescope noise data used in the training. The top panel shows a signal resulting from the merger of two black holes with masses $m_1^{src}=150\,M_{\odot}$~ and $m_2^{src}=110\,M_{\odot}$, generated at a luminosity distance of 40 Gpc using IMRPhenomPv2. The central panel shows the signal injected into the ET noise background. Note that this example is not taken from the MDC dataset.}\label{fig:ET-spectrogram-random}
\end{center}
\end{figure}

Having developed an understanding of how the model handles noise-only spectrograms, we now estimate its ability to distinguish anomalous input data. The ET MDC data set comprises a set of GW signals including 59,540 binary neutron star (BNS), 6,578 binary black hole (BBH), and 1,977 neutron star–black hole (NSBH) mergers. Here, we focus solely on BBH mergers, specifically those involving or forming IMBHs. The MDC data also provides spin, eccentricity and inclination parameters. We use the IMRPhenomPv2 approximant and include these in the generation of signal waveforms~\citep{PhysRevD.93.044007}. For signal-injected samples, the
merger time is required to lie within the spectrogram window,
but its position is randomly shifted within the segment (up to $\pm$30\% around the center of the window) in order to prevent
the model from learning a fixed time alignment and to reduce
overfitting.
Figure \ref{fig:ET-spectrogram-random} provides an example of the signal injection process, with spectrograms depicting the signal only (top, illustratively placed in the center of the spectrogram), noise only (bottom), and the full signal-injected example (middle).  

\begin{figure}[!ht]
\begin{center}

\resizebox{9cm}{!}{
\begin{tabular}{c}
\includegraphics{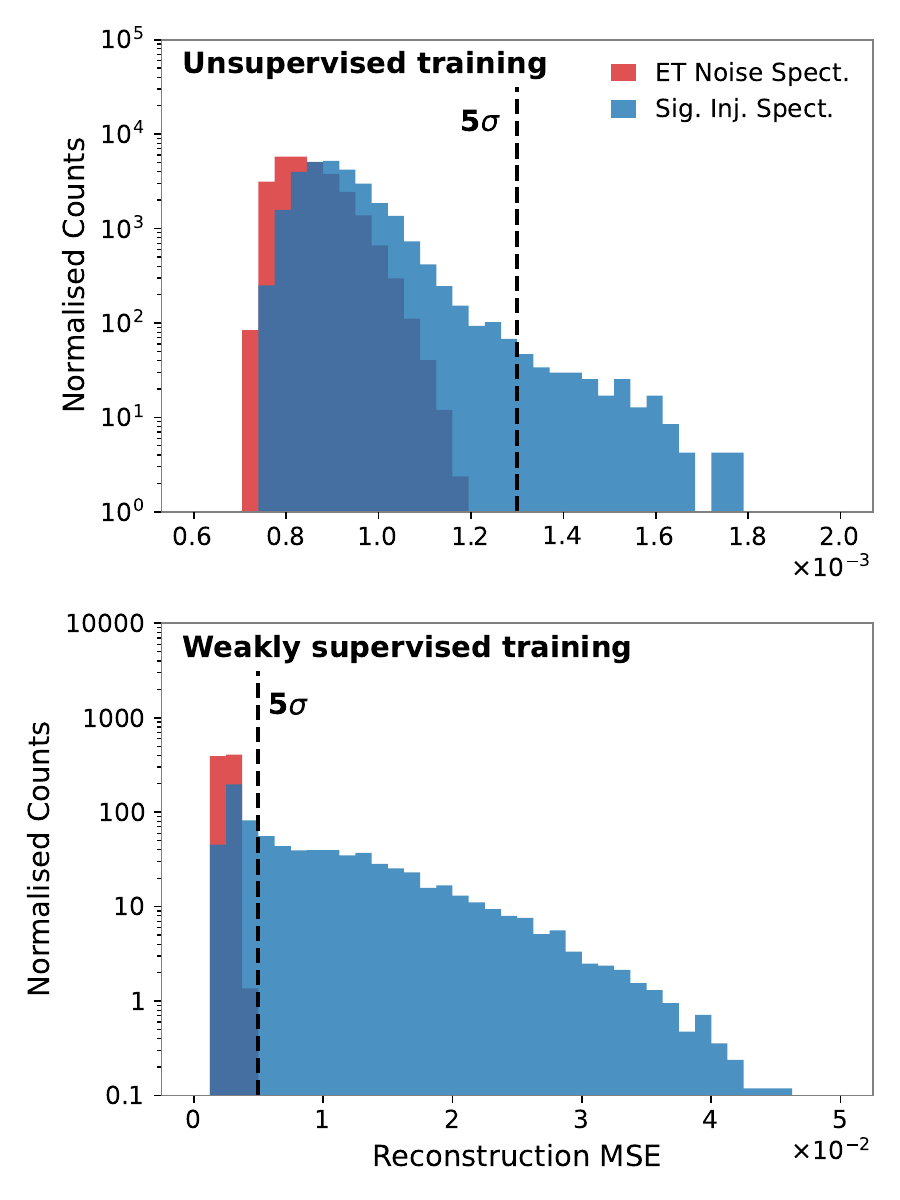}
\end{tabular}}\caption{The MSE values calculated for MDC samples injected into ET noise backgrounds. Shown in red is the distribution of 1$\times 10^{5}$ ET noise-only spectrograms. The results using the model trained noise-only spectrograms are shown on the top; the bottom plot shows instead the MSE calculated by the model, which is trained using the weak supervision method described previously.  In both, the calculated threshold value above which a spectrogram is flagged as anomalous is shown.}
\label{fig:SNR_vs_source_mass}
\end{center}
\end{figure}

In this section, we document comparisons between models developed with the basic unsupervised training and with the weak supervision method that is described previously. In each case, a detection threshold value of 5$\sigma$ is defined and applied as the criterion to flag an anomaly. Figure \ref{fig:SNR_vs_source_mass} contains two plots that show reconstruction MSE distributions calculated by the two models; the model unsupervised during training in the top figure, the weakly supervised model instead in the bottom. In both plots, the results from noise-only spectrograms are shown in red, while the full set of MDC binary black hole merger samples (spanning the full mass range) is shown in blue. 

\begin{figure}[!ht]
\begin{center}
\resizebox{9cm}{!}{
\begin{tabular}{c}
\includegraphics{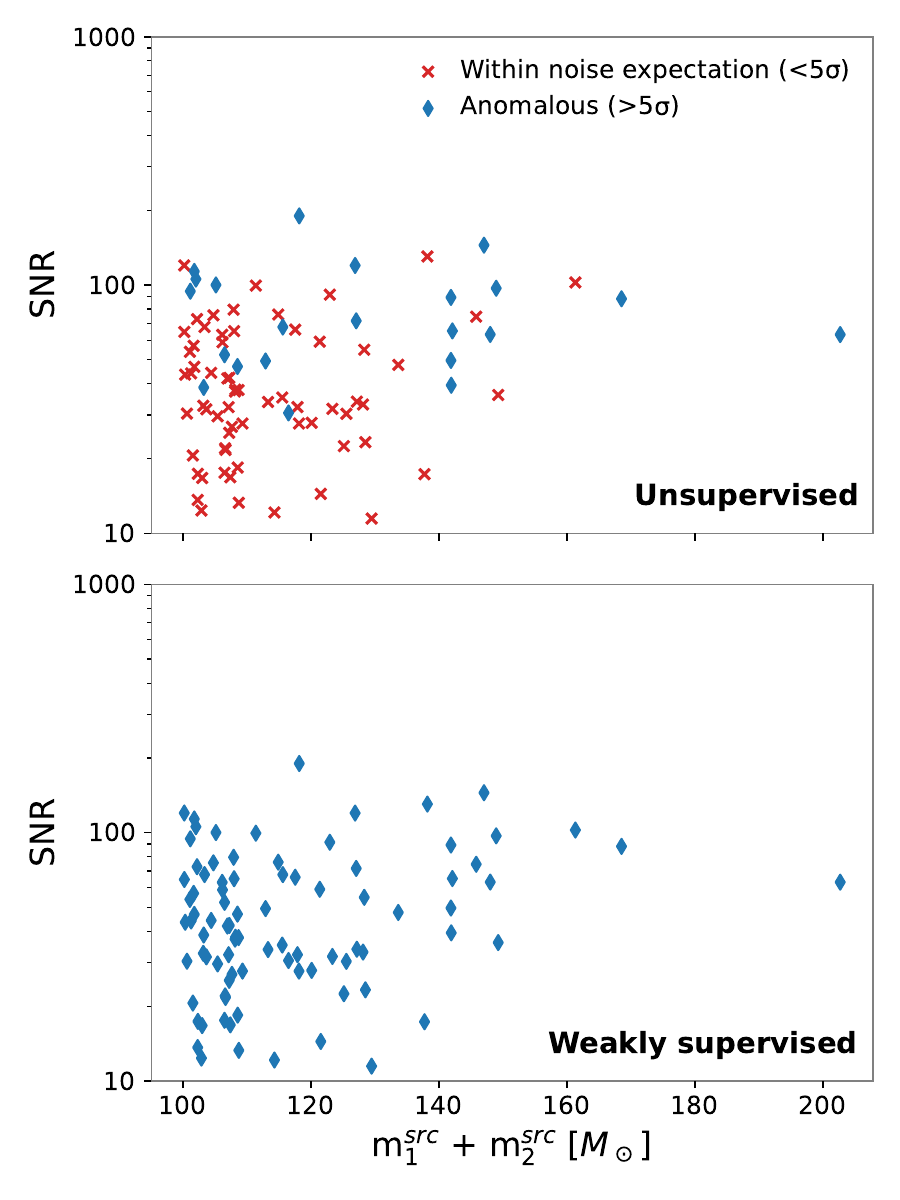}
\end{tabular}}
\caption{2D scatter plots of ET signal-to-noise ratio against total source frame mass of those mergers from the MDC sample with an IMBH expected to result from the merger. Those events which are flagged as anomalous by the classifier are shown with blue diamonds, those that are not are shown with red crosses. The top plot shows the results of using the model trained only on ET noise spectrograms, the bottom instead shows the results obtained with the model trained with the weak supervision described previously.}
\label{fig:IMBH_SNR_vs_source_mass}
\end{center}
\end{figure}

One can quickly see the significant improvement that can be gained with the implementation of the weak supervision during the training process. In both cases, there is a large portion of the injected signals that appear indistinguishable from noise; however, it is important to note that many such signals are likely outside of the target mass range of this study. Furthermore, the difference in scale between the two plots demonstrates the effect of the newly defined $m$ hyperparameter, which quantifies the target degree of MSE separation between noise and anomaly spectrograms. With the selected value of $m=5\times10^{-2}$ one can see the model has endeavored to push the distribution of reconstruction MSE towards such separation.

\begin{figure}[!ht]
\begin{center}
\resizebox{9cm}{!}{
\begin{tabular}{c}
\includegraphics{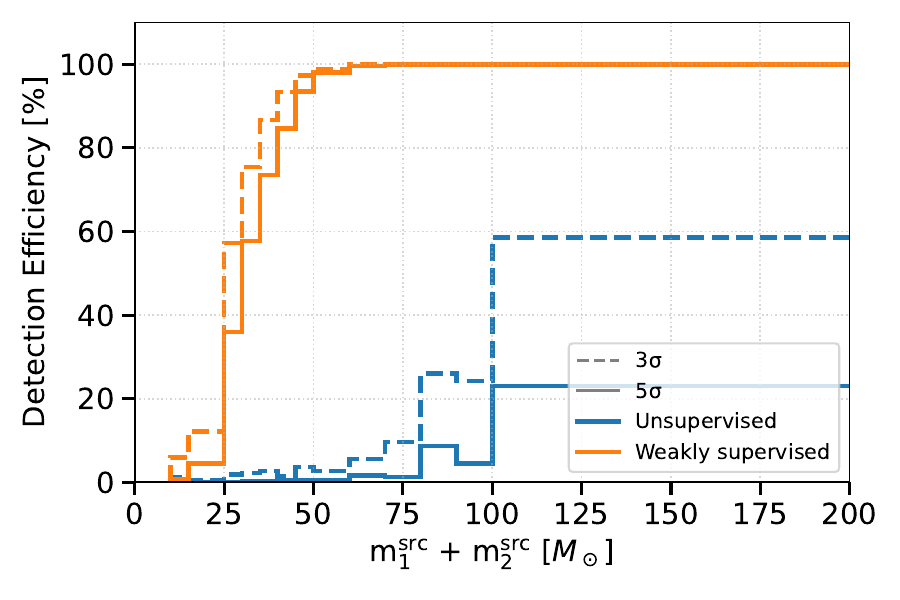}
\end{tabular}
}
\caption{Efficiency of signal recovery as a function of the total merger mass in the source frame. The efficiencies are calculated using either the 3 or 5$\sigma$ MSE thresholds. The result using an ET noise-only spectrogram model is shown in blue, and in orange is shown the result using a model trained by the described weak supervision method.}
\label{fig:eff_vs_mass}
\end{center}
\end{figure}

\begin{figure}[!ht]
\begin{center}
\resizebox{9cm}{!}{
\begin{tabular}{c}
\includegraphics{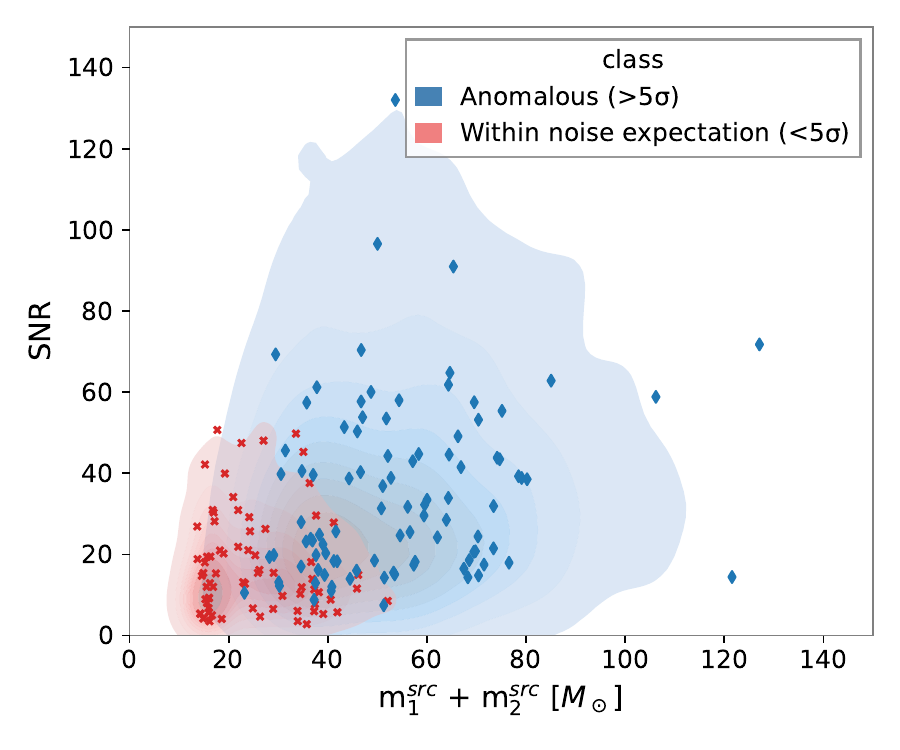}
\end{tabular}
}
\caption{Distribution of ET SNR against the sum of source frame masses, with successfully recovered signals shown in blue and those falling below the 5$\sigma$ threshold shown in red. This figure shows the kernel density estimation of the full MDC dataset which contains all 6578 BBH mergers. The scatter plots instead show just a random 2.5\% sample of this set.}
\label{fig:KDE_plot}
\end{center}
\end{figure}

Figure \ref{fig:IMBH_SNR_vs_source_mass} instead focuses directly on those mergers that lie within our target of IMBHs. Again, we show the results for both unsupervised (top) and weakly supervised (bottom) models, with the ET signal-to-noise ratio plotted against the total source frame mass in solar masses. We show individual mergers, denoted by red crosses when the model has failed to distinguish them from noise, and blue diamonds when the reconstruction MSE is greater than the 5$\sigma$ threshold. The improvement brought about by the weak supervision is clear, with 100\% of the mergers being recovered by the weakly supervised model. The unsupervised model, instead, is only able to recover 23\% of this same set of mergers. Figure~\ref{fig:eff_vs_mass} shows the total efficiency of the two models as a function of the total source frame mass. The results for the unsupervised and weakly supervised models are shown in blue and orange, respectively, with efficiencies resulting from the application of the 3 and 5$\sigma$ thresholds shown with dashed and solid lines, respectively. This demonstrates that the weakly supervised model is not only able to perform excellently in the targeted IMBH region, but is also able to generalise well to lower masses, with efficiency remaining at $\sim$100\% for total source masses as low as $M = 50\,M_\odot$.\\Figure~\ref{fig:KDE_plot} shows the kernel density estimation for the distribution of ET signal-to-noise ratio vs. source-frame masses for the full MDC dataset, containing all 6578 BBH mergers. Again, those events exceeding the 5$\sigma$ threshold are shown in blue, and those failing to be recovered are shown in red. A randomly selected 2.5\% sample of the mergers are shown in the scattered points. Mergers with SNR above approximately 40, or total source-frame mass exceeding about $30\,M_{\odot}$, are more often than not successfully recovered, with improving performance for higher masses. Conversely, merger events with a sum of source-frame masses below $20\,M_{\odot}$ are almost never recovered, even at higher SNR values. It is important to note that such low-mass mergers lie well outside the short-duration IMBH signal target of this study, and indeed, this could be indicative of a limitation in the 2-second spectrogram window adopted here. 

\section{Conclusions and Outlook}
\label{sec:conclusions}
In this paper, we demonstrated that anomaly detection algorithms based on simple convolutional autoencoders can be employed for the detection of gravitational waves, providing an alternative approach, for example, to waveform-dependent searches. We have shown that in realistic noise scenarios, such as simulated ET noise data, injected signals can be recovered with an efficiency of up to 100\%, for the cases studied, with a FAR due to possible statistical fluctuations of noise of ~4.6 events/year for a hypothetical interferometer collecting data with a duty cycle of 100\%. 

While the model was initially developed for possible IMBHs or bursts searches, we observe excellent generalisation capabilities and scalability. The model, in fact, after introducing weakly-supervised training, detects all mergers involving or forming IMBHs that are injected from the ET MDC. The training of the anomaly detection algorithm presented here is primarily based on learning from the expected noise distributions, with some weak supervision pushing for stronger separation of anomalous data. With further dedicated fine-tuning of parameters, this initial model shows promising ability in the detection of GWs originating from sources other than those considered in this study (\textit{i.e.} IMBHs).

The model, however, is not only sensitive to GW transients but also to anything that differs from pure noise. Glitches, for example, would also be flagged as anomalies by reproducing a large reconstruction error, introducing an additional problem of false alarms. Moreover, the presence of overlapping signals in ET data was not accounted for. This is a weakness of the model, and in general of anomaly detection algorithms: anomalies can be detected, but their origin (whether astrophysical, instrumental, environmental, statistical, etc.) not yet discerned. 

This last point, however, opens the way to further optimisation. In the current design, the algorithm is applied to only one detector; however, it can also be applied to several detectors in a network, thereby reducing the possibility of statistical fluctuations in the noise or glitches if the detectors are not co-located. Furthermore, it would be possible to introduce an additional module to classify anomalies as possible signals, as done, for example, in~\citep{Bini:2023gil}, and eventually also estimate their parameters~\citep{Dax:2021tsq,Papalini:2025exy,Lanchares:2025kpb}. A combination of multiple machine learning modules aiming at anomaly detection of signal candidates, signal candidates classification, and parameter estimation, in a single framework, will provide a fully automated analysis pipeline capable of detecting GWs and generating low-latency alerts.

\section*{Acknowledgements}
The authors wish to thank the Einstein Telescope Collaboration and in particular the coordinators of Division 10 of the Observational Science Board for providing the mock data used in this paper and the Information\&Technology office of the Marietta Blau Institute, for providing and maintaining the HEP-GPU hardware used for training, validation, and testing of the methods developed for this study.




\bibliographystyle{elsarticle-harv} 
\bibliography{example}






\appendix
\section{Ringdown frequency}\label{app:qnm_freq}
To relate the merger (ringdown) frequency to the total source--frame mass of the coalescing binary, we adopt the expression for the dominant quadrupolar quasinormal mode (QNM) of the final black hole~\citep{Berti:2005ys}. The frequency of the $(\ell,m,n)=(2,2,0)$ mode can be written as:
\begin{equation}
f_{220} = \frac{c^3}{2\pi G M_f}\,\alpha(a_f),
\end{equation}
where $M_f$ and $a_f$ are the mass and dimensionless spin of the remnant black hole, respectively, G is the gravitational constant, $c$ is the speed of light,
and $\alpha(a_f)$ is an empirical function determined from numerical–relativity fits:
\begin{equation}
\alpha(a_f) = 1.5251 - 1.1568(1-a_f)^{0.1292}.
\end{equation}
Assuming a typical final spin of $a_f = 0.7$, we obtain $\alpha(a_f) \simeq 0.536$~\citep{Keitel:2016krm,Healy:2014eua}.
The remnant mass is then estimated from the source--frame total mass $M_T$ by accounting for the radiated energy and cosmological redshift as
\begin{equation}
M_f = (1-\varepsilon)(1+z)\,M_T,
\end{equation}
with $\varepsilon = 0.07$ representing the fraction of mass--energy lost in the form of gravitational waves~\citep{Keitel:2016krm,Healy:2014eua}. Combining these expressions yields the working formula:
\begin{equation}
f_{220}(z, M_T) =
\frac{c^3}{2\pi G (1-\varepsilon)(1+z)M_T}\,\alpha(a_f).
\end{equation}
This relation provides an estimate of the characteristic
post–merger frequency, which scales approximately as $f \propto 1/[(1+z)M_T]$. It is clear that the heavier the system and the further it is located, the lower the value of $f_{220}$ of the event, which is typically, and specifically for circular or quasi-circular orbits, $f_{220}^{IMBH}<100$ Hz (for reference, $f_{220}^{GW190521}\approx 70$ Hz). It should be noted that higher-order modes will have a large impact to the waveforms of highly eccentric or highly mass-asymmetric mergers, but the mergers will anyway happen at low frequency.
\section{Implementation details}\label{app:implementation}

The initial development of the CAE model described herein was conducted using simple 2-second-long Gaussian-like white noise spectrograms generated with standard deviation of $\sigma=1\times10^{-21}$, using a standard short-time Fourier transform. A total of $2\times 10^{5}$ spectrograms are generated and split 70\%/30\% for training/testing purposes. BBH merger signals are generated with the IMRPhenomPv2 waveform approximant~\citep{PhysRevD.93.044007}, with only masses and distance considered as parameters, for simplicity. For signal-injected samples, the merger time is required to lie within the spectrogram window, but its position is randomly shifted within the segment (up to $\pm$30\% around the center of the window) in order to prevent the model from learning a fixed time alignment and to reduce overfitting. No additional optimisation of the time–frequency representation was performed.

\begin{figure}[!ht]
\begin{center}
\resizebox{9cm}{!}{
\includegraphics[width=0.5\textwidth,trim={0 15mm 0 25mm},clip]{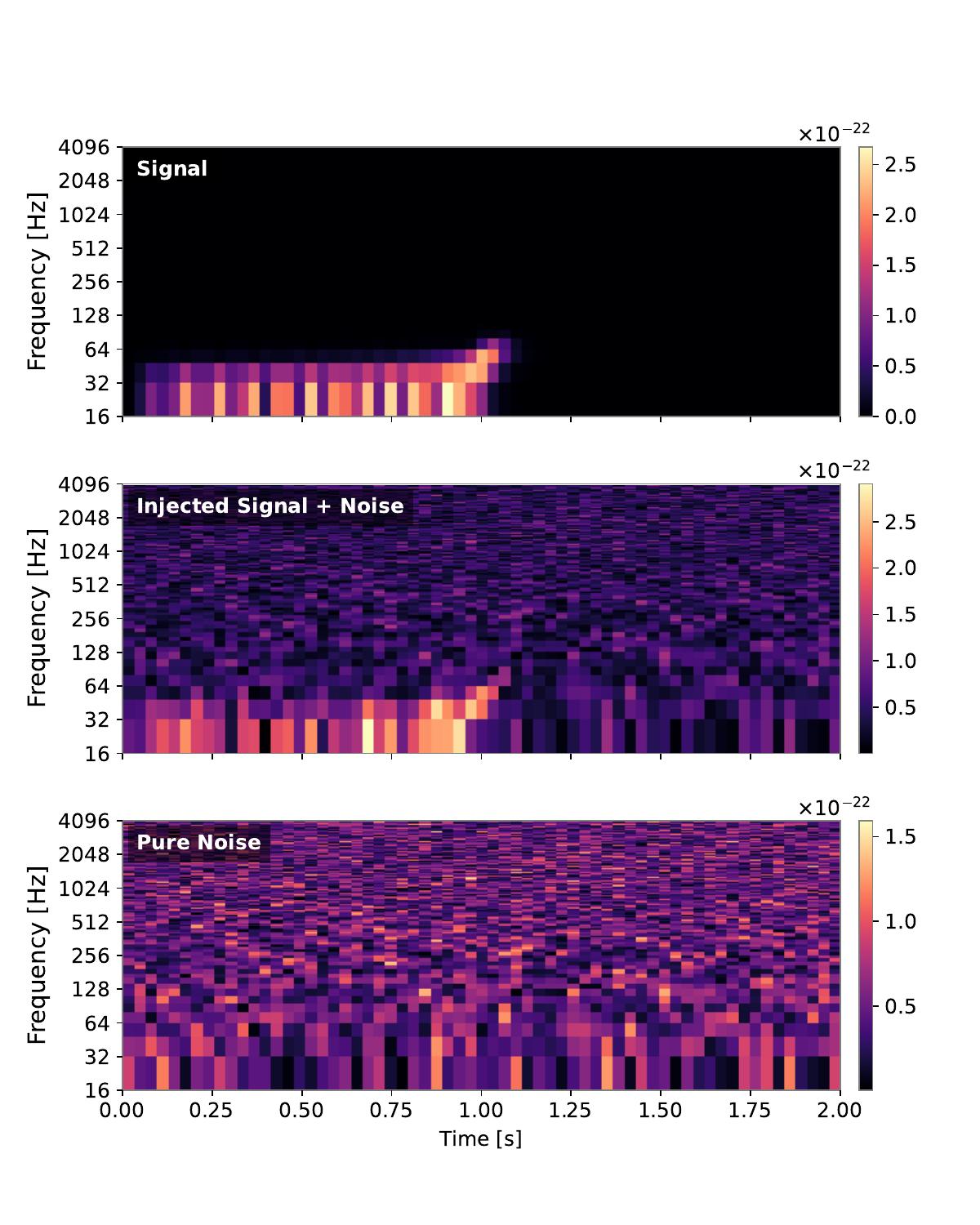}
}
\caption{Example white noise distributions used in the training and testing of the CAE. The lower panel shows a spectrogram of the pure white noise data used in the training. The top panel shows a signal resulting from the merger of two black holes with masses $m_1=150$ \msun~ and $m_2=110$ \msun, generated at a luminosity distance of 40 Gpc using IMRPhenomPv2. The central panel shows the signal injected into the ET noise background. Note that this example is not taken from the MDC dataset.}\label{fig:spectrograms-random}
\end{center}
\end{figure}

Although the IMBH signals studied herein inherently inhabit only the lowest frequency bins of the spectrograms, we elect not to apply any form of low-pass filters or cuts. This is done because, while this study specifically targets IMBH mergers, avoiding any such early selection criteria preserves flexibility for any future expansion of the method to a wider range of signal morphologies. Figure \ref{fig:spectrograms-random} shows spectrograms containing white-noise-only (bottom), generated signal (top, illustratively placed in the center of the spectrogram) and this signal injected into the noise (middle).

\begin{figure}[!ht]
\begin{center}
\resizebox{9cm}{!}{
\includegraphics{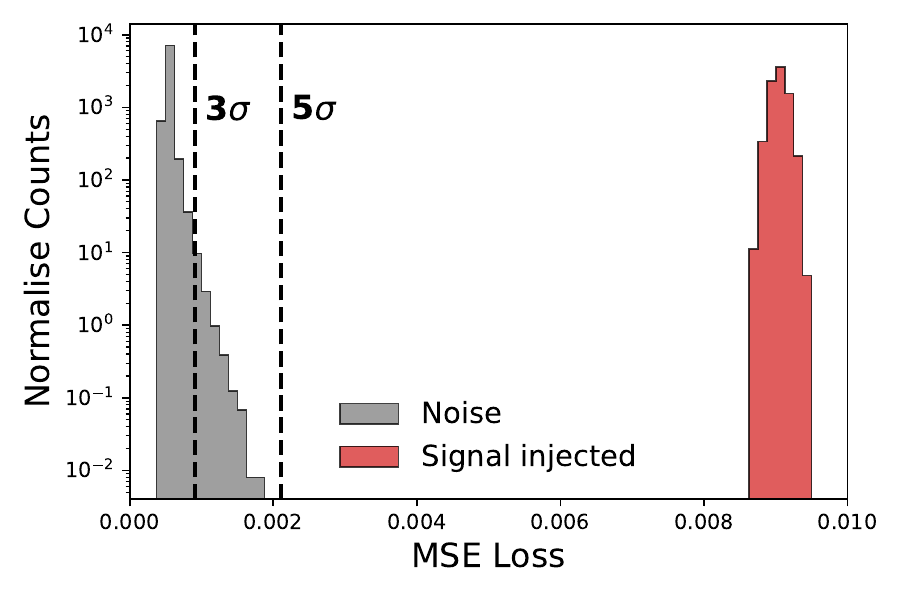}
}
\caption{Distributions of reconstruction MSE for white noise (grey) and signal injected (red) spectrograms. Here signals are situated at a fixed luminosity distance of 20 Gpc, with source masses sampled in the range 50 to 100 $M_\odot$.}\label{fig:whitenoise_separation}
\end{center}
\end{figure}

The initial use of white noise instead of ET simulated noise allows for simpler noise/signal separation and development of architecture and procedures. The 3 and 5$\sigma$ threshold values are shown, with the signal losses showing strong separation from white noise (Fig.~\ref{fig:whitenoise_separation}).

\end{document}